\documentclass[useAMS,usenatbib,usegraphicx]{mn2e}
\usepackage{natbib}

\title[X-ray lags in GX~339--4]{The evolution of the X-ray phase lags
  during the outbursts of the black hole candidate GX~339--4}

\author[Diego Altamirano and Mariano M\'endez]{Diego Altamirano $^{1}$\thanks{E-mail:
d.altamirano@soton.ac.uk} and Mariano M\'endez$^{2}$\\
$^{1}$Physics \& Astronomy, University of Southampton, Southampton, Hampshire SO17 1BJ, UK\\
$^{2}$Kapteyn Astronomical Institute, University of Groningen, PO Box 800, 9700 AV Groningen, The Netherlands.}

\begin{document}
%
%
%
\maketitle
\label{firstpage}
\begin{abstract}

Owing to the frequency and reproducibility of its outbursts, the
black-hole candidate GX~339--4 has become the standard against which
the outbursts of other black-hole candidate are matched up. 
Here we present the first systematic study of the evolution of the
X-ray lags of the broad-band variability component ($0.008 - 5$ Hz) in
GX 339$-$4 as a function of the position of the source in the
hardness-intensity diagram. The hard photons always lag the soft ones,
consistent with previous results.
In the low-hard state the lags correlate with X-ray intensity, and as
the source starts the transition to the intermediate/soft states, the
lags first increase faster, and then appear to reach a maximum,
although the exact evolution depends on the outburst and the energy
band used to calculate the lags.
The time of the maximum of the lags appears to coincide with a sudden
drop of the Optical/NIR flux, the fractional RMS amplitude of the
broadband component in the power spectrum, and the appearance of a
thermal component in the X-ray spectra, strongly suggesting that the
lags can be very useful to understand the physical changes that
GX~339--4 undergoes during an outburst. We find strong evidence for a
connection between the evolution of the cut-off energy of the hard
component in the energy spectrum and the phase lags, suggesting that
the average magnitude of the lags is correlated with the properties of
the corona/jet rather than those of the disc. Finally, we show that
the lags in GX 339--4 evolve in a similar manner to those of the
black-hole candidate Cygnus X--1, suggesting similar phenomena could be
observable in other black-hole systems.

\end{abstract}


%
%
%
%

\section{Introduction}\label{sec:intro}

The X-ray spectrum of black-hole candidate (BHC) systems in low-mass
X-ray binaries (LMXBs) can be decomposed into two main components:
\citep[e.g.,][]{Tanaka89,Vanderklis94}: a hard component, usually
fitted with a power law with photon index in the range 1.4--2.5
\citep[e.g.,][]{Thorne75, Sunyaev91, Grebenev93}, and a soft, thermal,
blackbody-like component with temperature $kT \sim 0.5 - 2$\,keV
\citep[e.g.,][]{Mitsuda84,Miyamoto93}. The hard component is usually
attributed to a corona of energetic electrons, and the state where it
dominates is known as the low-hard state (LHS).
The soft component is attributed to thermal emission from an optically
thick but geometrically-thin accretion disc \citep{Shakura73}; when
the soft component dominates the emission, the source is in the
so-called high-soft state (HSS).

Any description of the spectral evolution during an outburst based on
spectral fittings is of course model-dependent, therefore many authors
have opted for using X-ray colours (also known as hardness); these
colours are defined as the ratio between count rates in different
bands, and have proven to be very useful to understand the spectral
evolution of outbursts in terms of the dominating component of the
spectrum.
For BHCs, the hardness--intensity diagram (HID) is generally used to
characterize the spectral properties of a source. In what is now known
as ``canonical outburst'', the source evolves in a roughly square
pattern in the HID \citep[e.g.,][among many
  others]{Homan01,Belloni05}.  First, the intensity of the source
increases by a large factor while the hardness slightly decreases. The
source is in the LHS and traces a roughly vertical line in the HID.
At some point the source ``turns the corner'': the intensity stops
increasing and the source spectrum softens at more or less constant
intensity towards the HSS.
In between the LHS and HSS the source passes through the hard- and
soft-intermediate states \citep[see, e.g.,][]{Homan05}, showing
complex behaviour including sometimes large flares in intensity. After
the source reaches the HSS, the intensity usually decreases at
approximately constant hardness, with some excursions to harder or
softer states; as the outburst progresses, the source returns to the
LHS at approximately constant intensity. The hardening of the spectra
continues until the hard colour returns to a value similar to those
observed at the beginning of the outburst. At this point the hardening
stops, the intensity drops and the source returns to quiescence.

While the detailed evolution of BHC outbursts can be more complex than
what we have just described, change slightly from source to source, or
in different outbursts of the same source \citep[see, e.g.,][and
  references therein]{Homan01,Belloni05, Homan05, Remillard06, Vanderklis06},
the loop in the HID is usually recognizable.
This ``q-shaped'' loop is usually discussed in terms of hysteresis
\citep[e.g.][]{Miyamoto95}, and has been known for a long time
\citep[e.g.][]{Homan01,Maccarone03}. We note however, that although we
describe this hysteresis effect only in the case of BHs in LMXBs,
similar loops can be seen in HIDs and/or colour-colour diagrams of
other accreting objects \citep[e.g., neutron stars and Dwarf Novae;
  see e.g., ][]{Maccarone03, Kording08}, therefore pointing at a
similar physical origin.
The accretion physics behind the ``q'' shape in the HID of BHs, and
what produces the corners is not fully understood \citep[but see,
  e.g., proposed scenarios by][and references
  therein]{Meyer-Hofmeister05,Petrucci08,Begelman14,Kylafis15};
however, it is probably related to significant changes in the
geometrical/structural configuration of the system triggered by
changes in mass accretion rate ($\dot{M}$). For example, there is
clear evidence that the ejection of relativistic jets takes place
during some of the transitions between the states described above
\citep[e.g., ][]{Vadawale03,Fender04,Corbel04,Belloni07}.

\begin{figure} 
\centering
\resizebox{\columnwidth}{!}{\rotatebox{0}{\includegraphics[clip]{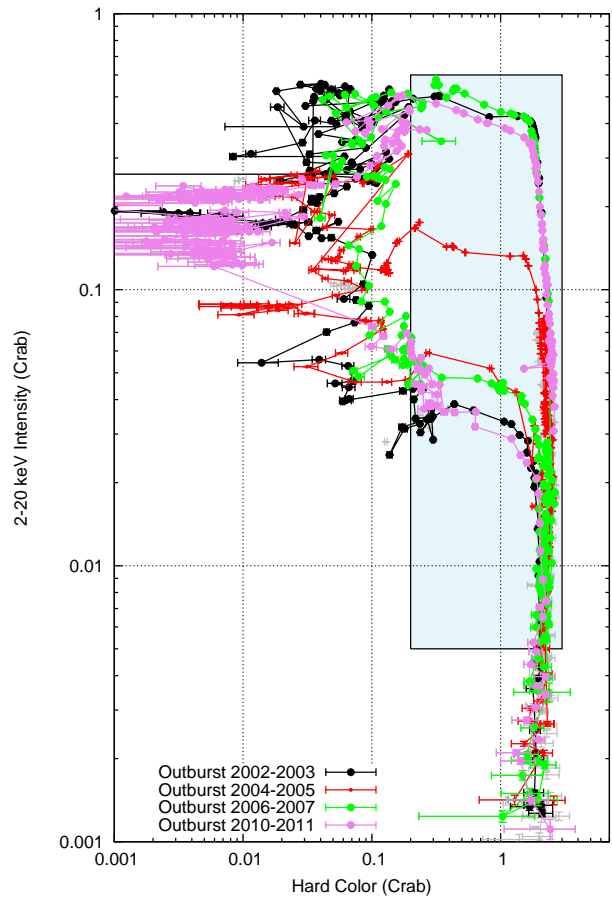}}}
\caption{Hardness-Intensity diagram of the four best sampled outbursts
  of the black hole GX~339--4. Hard colour is defined as the count
  rate in the 16.0--20.0 keV band divided by the count rate in the
  2.0--6.0 keV band. Intensity is the count rate in the 2.0--20 keV
  band. Both the hard colour and the intensity are normalized by the
  Crab (see Section~\ref{sec:dataanalysis} for more details).  The
  four different outbursts are plotted using different colours (see
  legend). Note that for the 2004-2005 outburst (red points) the
  hard-to-soft transition happened at a much lower (factor of $\sim$
  5) intensity than for the other three outbursts. For the other three
  outbursts the transition occurred at approximately the same
  intensity. The rectangular shaded area shows the part of the
  outbursts that we used in the rest of the paper (see
  Figure~\ref{fig:Lags}). The few gray points in this plot indicate
  observations in which the source was in the high-soft state, which
  we did not analyse in this paper (see
  Section~\ref{sec:dataanalysis}).  }
\label{fig:HID}
\end{figure}

To fully describe the evolution of a black-hole outburst it is
essential to consider the X-ray (time) variability. The power spectra
of the LHS and hard-intermediate state \citep[e.g.,][]{Mendez97a,
  Belloni05} are dominated by a strong band-limited noise component
that can reach fractional RMS amplitudes of up to $\sim50$\%
\citep[see][ and references therein]{Oda71,Mendez97a, Belloni97a,
  Vanderklis06}. Sometimes, low frequency quasi-periodic oscillations
(QPOs) are observed with frequencies in the range of
$\sim10^{-3}-20$~Hz. The characteristic frequencies \citep{Belloni02}
of these QPOs and noise components are found to correlate \citep[e.g.,
][]{Wijnands99a, Psaltis99b, Belloni02}, generally increasing towards
softer states.
The soft-intermediate state shows no strong band-limited noise
component \citep[e.g.,][]{Oda76,Homan01,Homan05}, and transient QPOs
appear with frequencies that are rather stable
\citep[e.g,][]{Casella04,Casella05}.
In the HSS, only weak power-law noise is observed in the power
spectrum and sometimes a QPO with frequency in the range 10--20~Hz
\citep[see, e.g., the HSS of XTE~J1550--564;][]{Homan01}.
Overall, the most common QPOs in BHCs have been classified based on
their characteristics as Type-A, B and C, and are usually associated
to different states \citep[e.g.,][]{Casella05}.

\begin{figure*} 
\centering
\resizebox{2\columnwidth}{!}{\rotatebox{0}{\includegraphics[clip]{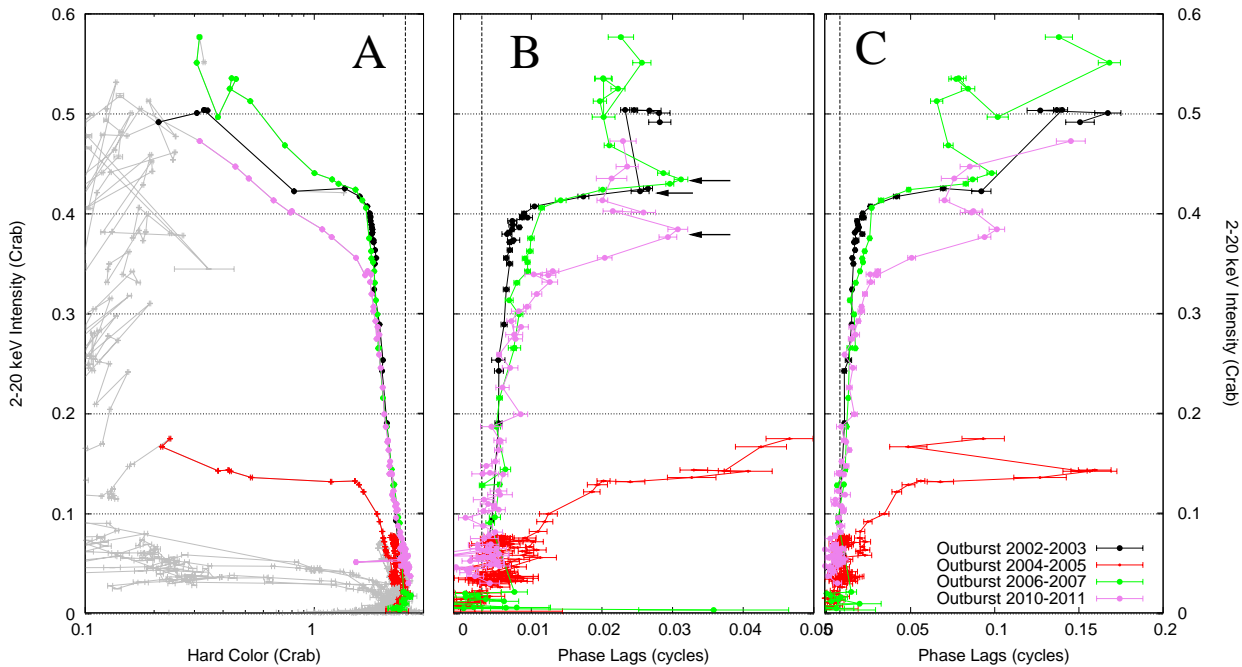}}}
\caption{Panel A shows a portion (see shaded area in
  Figure~\ref{fig:HID}) of the full hardness-intensity diagram (HID)
  of the four best sampled outbursts of the black hole GX~339--4 with
  RXTE.  Grey points indicate observations which were not studied in
  this paper (see Section~\ref{sec:results}). The black, red, green
  and violet points indicate the part of the four outburst (see text
  for details, and the legend for the identification of each outburst)
  for which we studied the lags. 
  Hard colour is defined as the count rate in the 16.0--20.0 keV band
  divided by the count rate in the 2.0--6.0 keV band. Intensity is the
  count rate in the 2.0--20 keV band. Both the hard colour and the
  intensity are normalized by the Crab (see
  Section~\ref{sec:dataanalysis} and Figure~\ref{fig:HID} for more
  details)
  The hard-to-soft transition during the 2004-2005 (red) outburst took
  place at an intensity that is a factor $\sim$5 lower than in the
  other three outbursts.
  Panel B and C show the phase lags-intensity diagram (LID) for the H1
  ($\simeq5.7-15$ keV) and H2 bands ($\simeq16-35$ keV),
  respectively. The lags correspond to the average phase lag divided
  by $2\pi$ in the $0.008-5$ Hz frequency range using as reference the
  soft band ($\simeq2-5.7$ keV). Horizontal arrows in Panel B mark the
  moment where the lags stop increasing with intensity (see
  Section~\ref{sec:results} for more details).
  The vertical dashed-lines show that, as GX~339--4 becomes brighter
  in each outburst, the spectrum becomes softer (panel A) and the lags
  increase (panel B \& C). The horizontal dotted-lines are shown to
  help the eye link the evolution of the lags as a function of the
  position of the source in the HID.
  In all panels, each point corresponds to the average value calculated
  from a full RXTE observation.}
\label{fig:Lags}
\end{figure*}

One important aspect of the variability of any astrophysical signal is
the possible phase lag between measurements in two different energy
bands \citep[e.g.,][]{Priedhorsky79,Nowak99}. The phase lag is a
Fourier-frequency-dependent measure of the phase delay between two
concurrent and correlated signals, in this case light curves of the
same source, in two different energy bands. The phase-lag spectrum of
the broad-band noise component in the power density spectrum of black
hole systems \citep{Miyamoto88,Nowak99} were first interpreted as due
to soft disc photons that are Compton up-scattered in a corona of hot
electrons that surrounds the system \citep[see, e.g.,][However, see,
  e.g., \citet{Nowak99, Maccarone00, Poutanen01, Arevalo06, Uttley11}
  for problems with this interpretation]{Thorne75, Hua97, Kazanas97}.
Phase lags from QPOs and the broad-band noise have been relatively
well studied in the LHS of a few sources, but using only a handful of
observations \citep[e.g][]{Vanderklis87,Nowak99, Wijnands99, Ford99,
  Cui00, Reig00a, Mendez13}.
If the geometry of the system changes significantly during an
outburst, one would naturally expect that the phase lags between soft
and hard photons will change as a function of the position of the
source in the HID.
To our knowledge, only the lags of the persistent (and somewhat
special) BH system Cygnus~X-1 have been studied in detail, taking
advantage of the full Rossi X-ray Timing Explorer (RXTE) archive. As
expected, the lags change as a function of the position of the source
in the HID \citep[see, e.g.,][and references therein]{Grinberg14}.

In this paper, we present the first multi-outburst study of the time
lags of the low-frequency variability for a transient BHC LMXB. For
this exploratory work, we chose the BHC GX~339--4, as evidence for lag
evolution has been reported for the 2002/2003 outburst \citep[see
  Figure 5 in][]{Belloni05}. GX~339--4 is probably one of the best
studied BHC sources today \citep[e.g.,][and references
  therein]{Mendez97a, Homan05a, Coriat09, Motta11a, Buxton12,
  Rahoui12, Corbel13}, and its outbursts are considered as
representative of the black hole population
\citep[e.g.,][]{Belloni05,Vanderklis06}.

\begin{figure*} 
\centering
\resizebox{2\columnwidth}{!}{\rotatebox{0}{\includegraphics[clip]{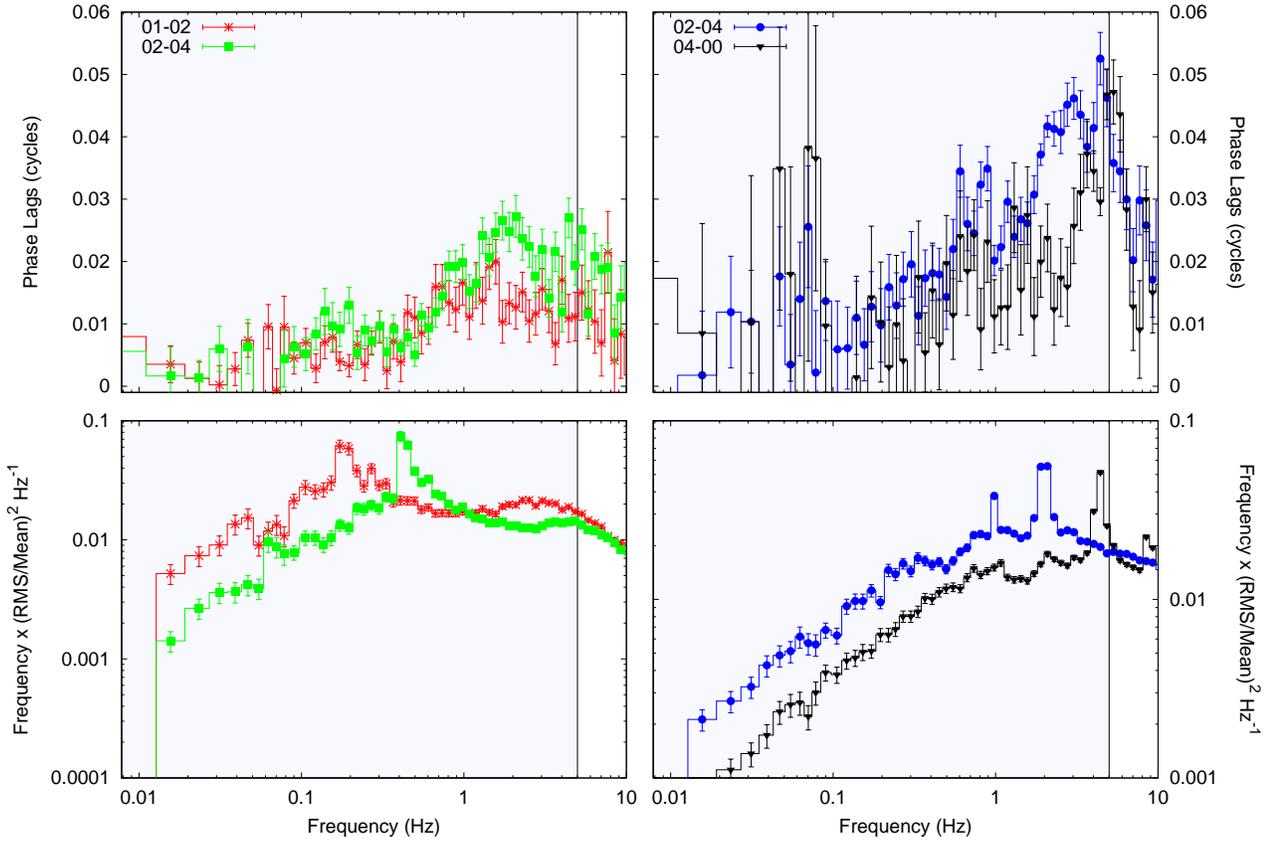}}}
\caption{H1 phase-lag (top; lags are in the $\simeq5.7-15$ keV band
  respect to the $\simeq2-5.7$ keV band. See
  Section~\ref{sec:dataanalysis} for more details) and power spectra
  (bottom; 2--60 keV) of four representative observations.  The red,
  green, blue and black colours mark observations 92035-01-01-02,
  92035-01-02-04, 92035-01-02-06 and 92428-01-04-00, respectively. The
  location of these observations in the HID and LID are indicated in
  Figure~\ref{fig:HIDPP}. }
\label{fig:PlagsPDS}
\end{figure*}

\begin{figure*} 
\centering
\resizebox{2\columnwidth}{!}{\rotatebox{0}{\includegraphics[clip]{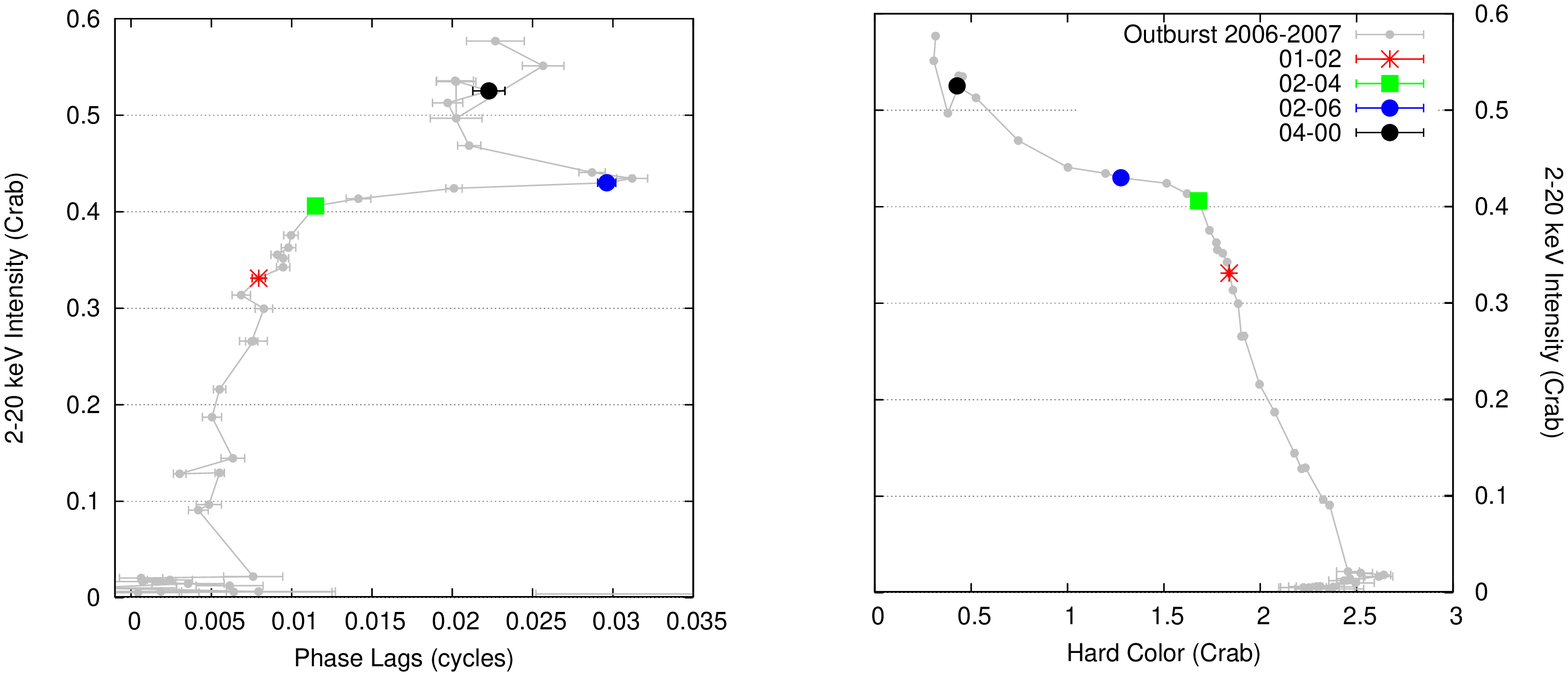}}}
\caption{Hardness-intensity diagram (right) and H1 Phase-Intensity
  diagram (left) for the 2006-2007 outburst.  Hard colour is defined
  as the count rate in the 16.0--20.0 keV band divided by the count
  rate in the 2.0--6.0 keV band. Intensity is the count rate in the
  2.0--20 keV band. Both the hard colour and the intensity are
  normalized by the Crab (see Section~\ref{sec:dataanalysis} for more
  details).
  The H1 phase-lags ($\simeq5.7-15$ keV) are calculated relative to
  the ($\simeq2-5.7$ keV) reference band, and correspond to the
  average phase lag divided by $2\pi$ in the $0.008-5$ Hz frequency
  range.
 The red, green, blue and black colours correspond to
  observations 92035-01-01-02, 92035-01-02-04, 92035-01-02-06 and
  92428-01-04-00, respectively. Their Phase spectrum and Power
  spectrum are shown in Figure~\ref{fig:PlagsPDS}. }
\label{fig:HIDPP}
\end{figure*}

\begin{figure} 
\centering
\resizebox{0.85\columnwidth}{!}{\rotatebox{-90}{\includegraphics[clip]{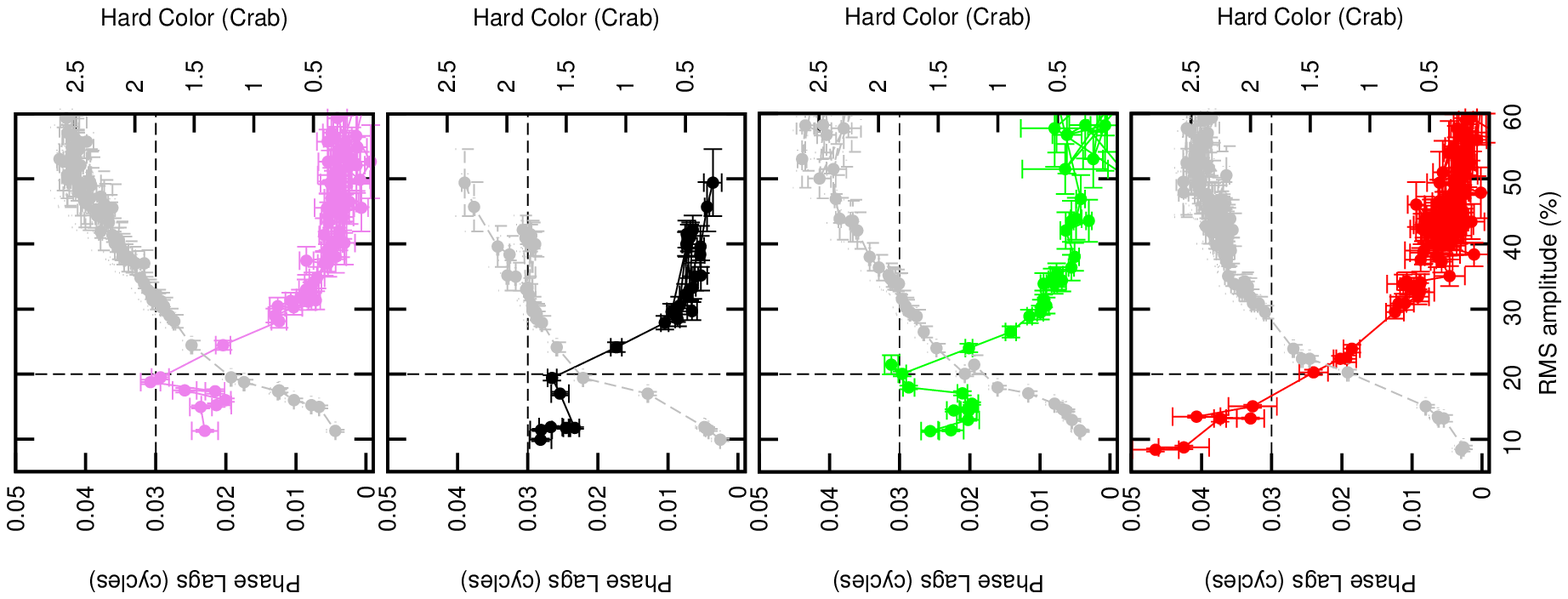}}}
\caption{$0.008-5$ Hz frequency average phase-lag divided by $2\pi$ of
  band H1 ($\simeq5.7-15$ keV) relative to the $\simeq2-5.7$ keV soft
  band (coloured points; left y axis) and hard colour (gray points;
  right y axis) vs. the average $0.008-5$ Hz fractional RMS amplitude
  (2--60 keV) for the four outbursts discussed in this paper (see
  Figure~\ref{fig:Lags} for the identification of the outbursts).
  Each point corresponds to the average values per observation, and
  for each measurement of a lag we plot the corresponding hard-colour
  point in gray. The hand-drawn dashed lines show the break in the
  hard-colour vs. RMS relation at RMS $\sim$20\%. The vertical and
  horizontal dashed lines are plotted to help the eye.}
\label{fig:LRH}
\end{figure}

%
%
%
%

\section{Observations and data analysis}\label{sec:dataanalysis}

We used all the 1414 pointed RXTE observations of GX~339--4 available
in the HEASARC archive. In particular for this study, we used data
taken with the Proportional Counter Array \citep[PCA;
][]{Zhang93,Jahoda06} over a time span of $\sim$14 years.
To calculate X-ray colours and intensity, we used the 16-s
time-resolution Standard 2 mode data and the procedure described in
\citet{Altamirano08}: for each of the five PCA (Proportional Counter
Units; PCU's) detectors we calculated a hard colour defined as the
count rate in the 16.0--20.0 keV band divided by the count rate in the
2.0--6.0 keV band. We defined the intensity as the count rate in the
2.0--20 keV band. Count rates in these exact energy ranges were
obtained by interpolating between PCU channels.
We corrected the data for dead-time, subtracted the background
contribution in each band using the standard bright source background
model for the PCA, version 2.1e\footnote{PCA Digest at
  http://heasarc.gsfc.nasa.gov/ for details of the model}, and removed
instrumental drop-outs. In order to correct for instrumental gain
changes \citep[see][]{Jahoda06}, we normalized all count rates by
those of the Crab Nebula, obtained in the same energy range, same gain
epoch and using the Crab observations closest in time to those of each
GX~339--4 observation. For each observation we then calculated colours
and intensity for each 16-s interval, and we averaged the colours and
intensity over all PCUs. Finally, we averaged the 16-s colours per
observation.

For the timing analysis we used data from the PCA Event, Good Xenon
and Single Bit modes. We calculated Leahy-normalized power density
spectra (PDS) using all photons in the PCA bandpass for data segments
of 128 seconds at the maximum time resolution available (except in the
case of Good Xenon mode, where we binned the data to a resolution of
1/8192-s). No background or deadtime corrections were applied to the
data prior to the calculation of the PDS. We subtracted a Poisson
noise spectrum that includes the effect of deadtime based on the
analytical function of \citet{Zhang95}, and then converted the
resulting PDS to squared fractional RMS \citep{Vanderklis95b}.

For each observation we also computed the complex Fourier transform
separately for all photons in the absolute channel bands $0-13$,
$14-35$ and $36-80$; in the rest of this paper we refer to these bands
as soft, H1 and H2, respectively.
These bands correspond to approximately the energy ranges $2-5.7$ keV,
$5.7-15$ keV and $15-34$ keV, respectively, and were selected in order
to maximize the number of observations where we could use the exact
same channel bands. While we used always the same channel-ranges, the
corresponding energy-ranges varied slightly in time due to the
instrumental gain changes \citep{Jahoda06}.
We estimated that those variations affect the lags by less than
$\sim 10$\%.
We calculated average cross spectra using the soft band as the
reference band, and following the description in \citet{Vaughan97} and
\citet{Nowak99}, we calculated the phase lags as a function of Fourier
frequency for each of the cross spectra.
We use the nomenclature H1 and H2 as well to refer to the lags between
photons in the H1 or H2 bands, respectively, relative to the photons
in the soft band.
Given our convention, positive lags means that hard X-ray photons lag
the soft ones.

A visual inspection of our data indicates that the shape of
  the phase-lag spectra is complex and dependent on the observation.
  One could divide each lag spectrum into different components, as it
  is commonly done for the power spectra of these same observations
  \citep[generally in the form of multi-Lorentzian model, see, e.g.,
  ][]{Nowak00,Belloni02}. However, the quality of the lag-spectra does not
  always allow to differentiate between components; therefore, as a
  first approach, in this paper we study the evolution of the
  frequency-averaged 0.008--5 Hz phase-lag\footnote{We calculated the
    frequency-averaged 0.008--5 Hz phase-lag, by first averaging the
    raw cross-correlation vectors, then averaging over frequency, and
    finally calculating the lag as $ PL = \arctan{(\Im/\Re)} / 2\pi$,
    where $\Im$ and $\Re$ are the imaginary and real part of the
    resulting frequency-averaged vector, respectively.}.
 This type of analysis is analogous to those works where the relation
 between average fractional rms amplitude of the power spectra of BHs
 (and also NS) is studied as a function of the spectral state of a
 source \citep[e.g. ][]{Belloni05,Fender09,
   Motta11a,Munoz11,Stiele11}. A possible multi-component description
 of the phase-lag spectrum will be discussed in a separate paper.

Dead-time driven cross talk between different energy channels induces
a $\pm \pi$ phase lag in the Poisson-noise dominated part of the PDS
\citep{Vanderklis87}. To correct for this, we subtracted the cross
spectrum in a frequency range well above that of the broad-band noise
($200-1000$ Hz), in which the X-ray variability is dominated by
Poisson noise. We inspected and confirmed that, as expected, the cross
spectrum in this frequency range has no significant imaginary part;
hence this procedure has no effect on the sign of the phase lags
\citep[see][for more details]{Vanderklis87}. Unless stated explicitly,
quoted errors use $\Delta\chi^2 = 1.0$.

Throughout the paper we give phase lags in units of
$\mathrm{rad}/2\pi$ (cycles) and we do not report on time-lags
(defined as phase-lag divided by the frequency $\nu$), as the
frequency range we use (0.008-5 Hz) spans nearly a factor $\sim$600 so
the average time-lag will strongly depend on the binning used
(e.g. linear or logarithmic), and can be strongly biased by the lags
at frequencies lower than 1~Hz. The average phase-lags measurements we
report here can be used to calculate time-lags at a given frequency to
a first-order approximation.

%
%
%
%

\section{Results}\label{sec:results}

We were able to measure the phase-lags in the 3 bands defined in
Section~\ref{sec:dataanalysis} in 1362 out of the 1414 observations
available in the archive. Approximately 1020 out of these 1362
observations sample four complete outbursts of GX~339--4 that showed
the canonical full loop in the HID; the HIDs of the four outbursts
studied here are plotted in Figure~\ref{fig:HID}.
As described in Section~\ref{sec:intro} \citep[and in previous works,
  e.g.,][and references therein]{Belloni05,Motta11a} and shown in
Figure~\ref{fig:HID}, the outbursts of GX~339-4 evolve in an
anti-clockwise ``q-shaped'' form.

In Panel A of Figure~\ref{fig:Lags} we show a portion of the HID of
these four outbursts. For each outburst the gray points show the full
outburst, while we indicate the rising part of each outburst (hard
colour $>0.2$) with the black, red, green and violet points for the
2002-2003, 2004-2005, 2006-2007, an 2010-2011 outbursts,
respectively. The transitions between the hard and the soft state are
well sampled with RXTE observations, except for the 2002-2003 outburst
(black), in which the hard-to-soft transition was sampled by only 5-6
observations.
In panels B and C of Figure~\ref{fig:Lags} we plot the source
intensity versus the 0.008-5 Hz phase-lags for the H1 and H2 bands,
respectively; in both cases we use the soft band as reference. We will
refer to these plots as the (Phase-)Lag-Intensity Diagrams, LID.

For the 3 brightest outbursts, Panel B in Figure~\ref{fig:Lags} shows
that the evolution of the H1 lags can be divided into 3 intervals:
initially the lags show a slow, but significant, increase with
intensity until the source ``turns the corner'' in the HID; at that
point the lags start to increase at a much higher rate with
intensity. Finally, the lags appear to reach a maximum at $\sim$0.03
cycles , decrease to $0.020-0.025$ cycles, and remain
approximately constant with intensity. After this point we can no
longer measure the lags in single observations, or the source starts
undergoing fast state transitions.  For the sake of clarity, this more
complex behaviour will be presented in a followup paper.

For the 2002-2003 (black), 2006-2007 (green) and 2010-2011 (violet)
outbursts, the H1 lags reach the maximum value on MJD 52400-52402
(ObsID: 70109-01-06-00/10-00), MJD 54140.2 (ObsID: 92035-01-03-00 and
MJD 55298.7 (ObsID: 95409-01-14-03), respectively (marked with
horizontal arrows in Figure~\ref{fig:Lags}, Panel B).  These dates
correspond to the moment when the Type-C QPO reaches frequencies in
the $2-3$ Hz range, close to, but still below, our 5-Hz limit for the
calculation of the lags. To investigate whether this could be the
reason for the behaviour of the lags at this point in the outburst, we
also calculated the average phase-lag in the $0.008-10$ Hz range (not
shown); in this case we found that the overall pattern in the LIDs is
the same as those shown in Panel B in Figure~\ref{fig:Lags}, with a
maximum of $\sim$0.03 cycles happening at the same position of the
arrows in Panel B.  In none of the observations of the 3 brightest
outbursts we detected a Type-B QPO; these QPOs occur when the source
softens further in the outburst \citep[see, e.g., table 3
  in][]{Motta11a}.

In the case of the 2004-2005 outburst (red), the lags increase
with intensity during the whole rising part of the outburst. The
``turn the corner'' point occurs at an intensity of $\sim$0.12 Crab,
and the softening in the HID is less sharp than in the other 3
outbursts.  This might be connected to the fact that this is also the
only full outburst which made the hard-to-soft transition at a
significantly lower luminosity than the others.
In this outburst the lags in the H1 band are the longest (up-to
$\sim$0.045 cycles). The last two observations presented here
(ObsID:90110-02-01-03/02-00), the one with the longest lags, already
contain Type-B QPOs in the power spectrum.

The evolution of the H2 lags during the individual outbursts in Panel
C is similar to what we observe in Panel B, with the main differences
being that: (i) the lags during the 2004-2005 outburst (red) show a
maximum of $\sim$0.17 cycles (at MJD 53232.35,
ObsID:90110-02-01-00/02), and a sudden drop after that; the two
observations after the maximum in this Panel correspond to those where
we detect Type-B QPOs (see above), and (ii) the lags measured in the
other 3 outbursts show a fast increase at the very end of our
sampling, reaching a maximum in the $\sim$0.14--0.17 cycles range.

In all outbursts, the point where the lags start increasing at a
higher rate (see panels B \& C in Figure~\ref{fig:Lags}) coincides
with the ``turn the corner'' point in the HID (see panel A), where the
source starts the rapid transition to the intermediate state (in all
panels of Figure~\ref{fig:Lags} we plot horizontal dotted-lines to
guide the eye), strongly suggesting that both phenomena are connected,
and related to the same changes the system is undergoing during this
softening.
This ``turn the corner'' point in the HID does not appear to be
related to the appearance of the Type-C QPO, as at least in the
2006-2007 and 2010-2011, clear Type-C QPOs are already detected 
before this transition \citep[e.g.,][]{Motta11a}.

In Figure~\ref{fig:LRH} we plot the H1 lags (left y axis) vs. the
fractional RMS amplitude computed in the $0.008-5$ Hz range, using
the full 2--60 keV RXTE energy band (coloured points). To compare with
the correlation between fractional RMS amplitude and hard colour
\citep[e.g.,][]{Belloni05}, we also plot in this Figure the relation between 
hard colour (right y axis) vs. fractional RMS amplitude (gray points).
For the 3 brightest outbursts, the point where the H1 lags reach
their maxima (panel B of Figure~\ref{fig:Lags}) corresponds to the point 
where the average RMS amplitude is $\sim20$\%. Again, this is not the 
case for the 2004-2005 (red) outburst, where the H1 lags continue 
increasing up to the end of our sampling.
From Figure~\ref{fig:LRH} it is also apparent that the hard-colour
vs. RMS relation has a break when the RMS amplitude is $\sim$20\% (see
hand-drawn dashed lines in each panel), even for the 2004-2005
outburst.  \citep[This break was already apparent, but not
  specifically discussed, in some of the Figures in][]{Belloni05,
  Fender09, Motta11a, Stiele11}.

We note that GX~339--4 has undergone additional outbursts
\citep[e.g,][]{Motta11a}, however we do not report them here in detail
as they were either outbursts where GX~339--4 always remained in the
hard state, or the outburst was not well sampled by RXTE
observations. Nevertheless, the lags during the hard state of
those other outbursts are consistent with the ones shown in
Figure~\ref{fig:Lags}.
In Figure~\ref{fig:Lags}, when hard colour $<0.2$, the RMS amplitude
in the PDS decreases further \citep[not shown in Figure~\ref{fig:LRH};
  see e.g.,][]{Motta11a} and, as mentioned above, the lags in
single observations are not always well constrained.
During the soft-to-hard transition at the end of these outbursts, the
RMS amplitude in the PDS increases once more
\citep[e.g.,][]{Belloni05, Stiele11}, and we can again measure the
lags in individual observations; in this phase of the outburst
the lags are consistent with those in the LHS during the rise of
the outburst. We do not discuss those lags further in this paper.

%
%
%


\section{Discussion}

We present the first systematic study of the evolution of the X-ray
lags of the broad-band variability component ($0.008 - 5$ Hz) in the
black-hole candidate GX 339$-$4 as a function of the position of the
source in the hardness-intensity diagram (HID). The hard photons
always lag the soft ones, with the phase lags ranging from $\sim
0.005$ to $\sim 0.17$ cycles depending on the observation
and the energy band used to measure the lags. In all four outbursts
studied here, as the source brightens at the beginning of the
outburst, the lags initially increase slowly as the source is in the
low-hard state (LHS), and then they start to increase much faster with
intensity as the source initiates the transition to the
hard-intermediate state (HIMS). After reaching a local maximum, the
lags decrease and remain more or less constant
(Figure~\ref{fig:Lags}B), or decrease and then increase again
(Figure~\ref{fig:Lags}C), before the source moves into the high-soft
state (HSS). While there is no unusual feature in the HID at the time
that the lags reach the maximum value, the maximum of the lags
coincide with a significant break in the relation between the
hard-colour and the RMS amplitude of the broad-band component in the
power spectrum of the source. For the 2002-2003 outburst, this point
also coincides with the time at which \citet{Homan05a} observed a
dramatic drop of the optical/NIR flux which they interpreted as a
sudden change in the properties of the compact jet in this source.

\begin{figure*} 
\centering
\resizebox{2\columnwidth}{!}{\rotatebox{-90}{\includegraphics[clip]{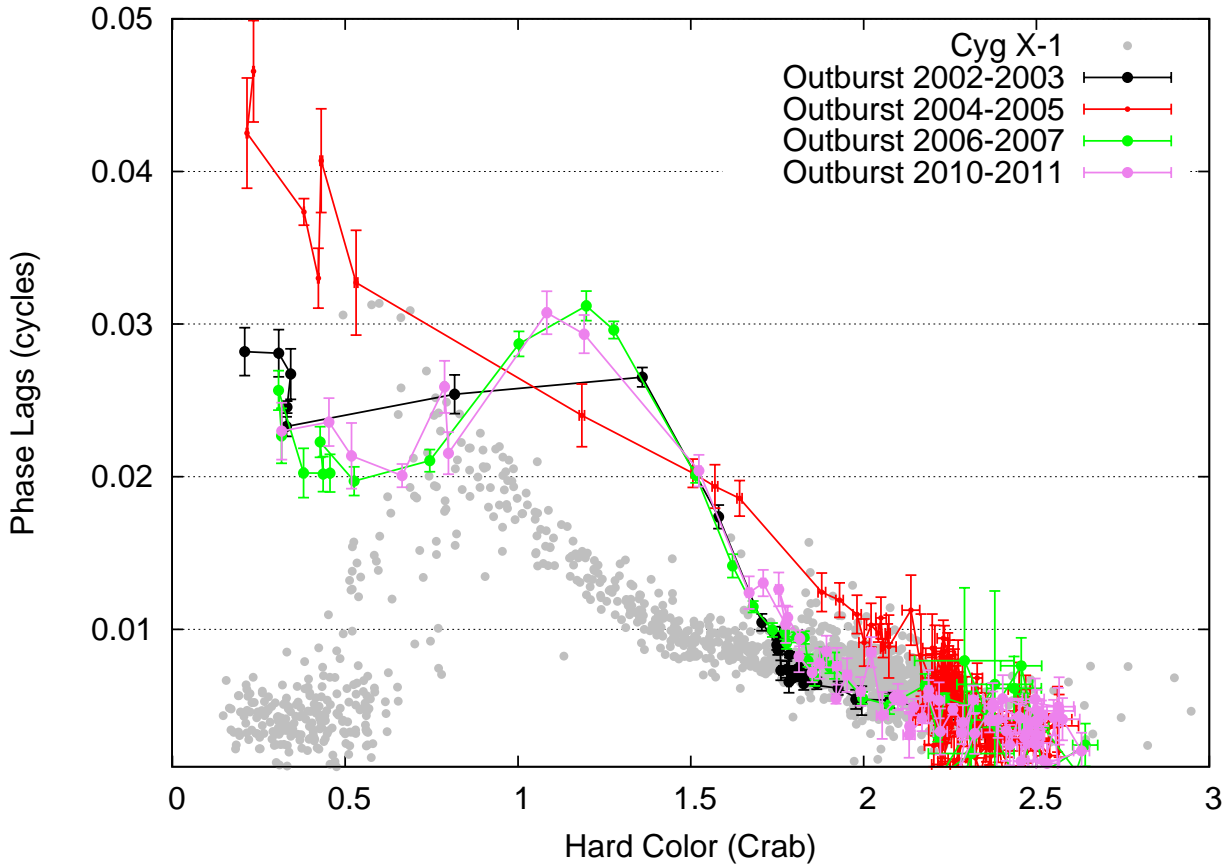}}}
\caption{H1 phase-lag ($\simeq5.7-15$ keV band respect to the
  $\simeq2-5.7$ keV band) vs. hard colour (normalized by the Crab) for
  Cyg X-1 (gray) and GX~339--4.  Hard colour is defined as the count
  rate in the 16.0--20.0 keV band divided by the count rate in the
  2.0--6.0 keV band. }
\label{fig:Comp}
\end{figure*}


\subsection{Phase-Lags and the Optical/NIR behaviour of GX~339--4}

Several works have studied the relation between the X-ray emission and
that at other wavelengths in Black Hole binaries \citep[see,
  e.g.,][and references therein]{Jain01a, Russell11, Buxton12}. In
particular, \citet{Homan05a} studied the X-ray and
Optical/Near-Infrared evolution of the 2002-2003 outburst of GX~339--4
(black points in our Figures~\ref{fig:Lags} and \ref{fig:LRH}). They
found that in the LHS (MJD 52330--52398), both the optical/NIR and
X-ray fluxes increase and remain well correlated with each other.  As
GX~339--4 evolved into the intermediate state, starting on MJD 52400,
the optical/NIR fluxes decreased rapidly until GX~339--4 reached the
spectrally soft state (about 10 days later), where the optical/NIR
fluxes remained low and approximately constant.
\citet{Homan05a} noticed that during the intermediate state, and
particularly on MJD 52402, the disc flux increased dramatically (from
$<1$\% to about 30\% of the total flux; note however that the total
flux in the RXTE band actually decreased slightly at this point)
without substantially changing nor interrupting the evolution of the
frequency of the Type-C QPO or the power-law index in the X-ray
spectra.
Unfortunately there is a 3-day gap, between MJD 52402 and 52405, in
the RXTE coverage of the 2002-2003 outburst (black points in
Figures~\ref{fig:Lags} and \ref{fig:LRH}).  If the lags in this
outburst evolved in the same way as in the 2006-2007 (green) outburst,
where the H1 lags drop from a local maximum of $\sim$0.03 cycles
to $\sim$0.02 cycles within one day, we would have probably
observed a maximum in the H1 lags in panel B (and possibly also in the
H2 lags, panel C) within a day or two of MJD 52402.
If that was the case, this would indicate that in GX~339--4 there is a
relation between the lags and the optical/NIR emission, whereas the
latter is not related to the power-law index or frequency of the
Type-C QPO \citep{Homan05a}. The RMS amplitude of the variability
would also be related to the optical/NIR emission, as there is a clear
break at $20$\% RMS amplitude in the hard colour vs. RMS relation,
coincident with the point where we observe the maximum in the H1 lags
(Figure~\ref{fig:LRH}; see below).

The accelerated steepening of the power-law component in the X-ray
spectrum around MJD 52400, together with GX~339--4 showing a dramatic
decrease in Optical/NIR fluxes, led \citet{Homan05a} to suggest that
this is the moment where the characteristics of the compact jet
\citep[as usually observed in the hard state, e.g.,][]{Corbel13} start
to change.
If the corona of hot Comptonizing electrons and the compact radio jet
are related, or if the base of the jet is the place where
Comptonization takes place \citep[e.g.,][]{Giannios04, Giannios05,
  Markoff05}, one would expect that the properties of the jet and the
lags are correlated. Indeed, given that in BHCs in the LHS the radio
and X-ray flux are correlated \citep[also for GX~339--4, e.g.,][and
  references therein]{Corbel13}, and that the lags correlate with
X-ray intensity (Panel B \& C in Figure~\ref{fig:Lags}), indicates
that, at least in the LHS, the lags probably correlate with the radio
flux of the compact jet.
Several spectral and timing properties of accreting systems correlate
with one another; here we have shown that the phase lags also
correlate with those properties. To progress further, and to gain
insight on what are the basic source properties that drive these
correlations, one would need to test, for instance, whether the lags
in GX 339--4 (and other sources) also correlate with the radio flux in
the intermediate state, where the lags do not (always) correlate with
the source intensity (see, e.g., Figure~\ref{fig:Lags}). Such a study
could prove whether the lags as we estimated in this work are a good
indicator of some of the radio jet characteristics, similar to the
evidence provided by the correlations between radio luminosity and
X-ray timing frequencies found in both NSs and BHs \citep[see,
  e.g.,][]{Migliari05}.


\subsection{Lags and the fractional RMS amplitude}

In Figure~\ref{fig:LRH} we showed that there is a break in the
hard-colour vs. RMS correlation that occurs consistently at RMS $\sim
20$\% in all four outbursts.
This break can also be seen as a ``notch'' in the absolute
RMS-intensity diagram \citep[e.g., Figure 1 in][at absolute RMS $\sim
  220$ cts/s and PCU2 count rate $\sim 1000$ cts/s, corresponding to a
  fractional RMS amplitude of $\sim 20$\%]{Munoz11}.
\citet{Munoz11} show that, at least in the 2006-2007 outburst, this
notch in the absolute RMS-intensity diagram coincides with the
observations where X-ray spectral fits require a thermal blackbody
component \citep{Motta11a}; \citet{Munoz11} interpreted this as the
appearance (in the 2--15 keV band) of an optically thick accretion
disc with a very low variability level.
For the 2002-2003 (black) outburst, the break in the hard-colour
vs. RMS relation also coincides with the time when the disc flux
showed a dramatic increase (from $<1$\% to about 30\% of the total
flux), and the optical and NIR bands started to decrease \citep[see
  discussion above, and ][]{Homan05a}.
As it can be seen in Figure~\ref{fig:LRH}, except in the case of the
weak 2004-2005 (red) outburst (see Section~\ref{sec:results} and
Figure~\ref{fig:LRH} for details), in the other three outbursts the
break in the hard-colour vs. RMS relation coincides with the moment
when the H1 lags reach their maximum, again suggesting a strong
relation between lags, RMS amplitude, and spectral evolution.

As it is apparent from Figure~5 in \citet{Fender09}, at least two
other BHC systems (XTE~J1550--564 and XTE J1859+226) also show a break
in the hard-colour vs. RMS relation that, coincidentally, happens at
around 20\% RMS as well. Although this coincidence has to be explored
in more detail \citep[as the frequency and energy bands used
  by][differ from those used in this paper]{Fender09}, it suggests the
interesting possibility that we can identify the point where the
spectral fits require a thermal component just by the use of the
fractional RMS amplitude or the lags, assuming that other sources
show the same evolution of the lags during the outburst.


\begin{figure} 
\centering
\resizebox{1\columnwidth}{!}{\rotatebox{0}{\includegraphics[clip]{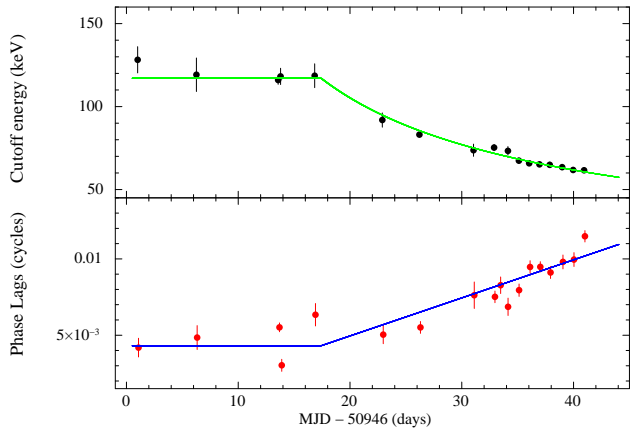}}}
\caption{Cut-off energy $E_{cut}$ of the energy spectra (top) as
  reported by \citet{Motta09} and H1 phase-lags (bottom; lags are in
  the $\simeq5.7-15$ keV band respect to the $\simeq2-5.7$ keV
  band. See Section~\ref{sec:dataanalysis} for more details) during
  the first 45 days of the 2006-2007 outburst of GX 339-4. With
  continuous lines we show a broken power law fit done to both curves
  simultaneously, with the condition that break is the same for both
  fits and that the power-law index before the break is zero. For this
  model, the break is at $MJD-50946=20.1\pm1$ day (see
  Section~\ref{sec:origin} for more details).}
\label{fig:cutoff}
\end{figure}

\subsection{GX~339--4 vs. Cygnus X--1}\label{sec:cygx1}

\citet{Grinberg14} reported on the long term evolution of the
energy-resolved X-ray variability of the persistent BHC LMXB
Cygnus~X--1 using data taken between 1999 and 2011.
These authors find positive time lags that evolve as a function of the
spectral state of the source: the average time lags (as calculated in
narrow frequency ranges) increase considerably when the source moves
from the hard to the intermediate state \citep[with a maximum of
  $\sim$20 msec for the 0.1--30 Hz lags of the 9.4--15 keV photons
  relative to the 2.1-4.5 keV photons; bands 4 and 1, respectively in
][]{Grinberg14}, and then the lags suddenly drop to $\la5$ msec when
Cygnus~X--1 enters the softest states.
To compare our results with those of \citet{Grinberg14}, we calculated
the average 0.008-5 Hz phase-lag for Cyg X-1 using the same energy
bands and the exact procedure used for GX~339--4. In
Figure~\ref{fig:Comp} we show the phase Lags of Cyg X-1 (gray
points\footnote{Note that the datapoints form a pattern very similar
  to those showed by \citet{Grinberg14}, although we are plotting an
  averaged phase-lag in the 0.008-5 Hz frequency range, while
  \citet{Grinberg14} were plotting average time-lag in narrow
  frequency ranges.}) and GX 339--4 (colour points; colours are the
same as in Figure~\ref{fig:Lags}).
This Figure  shows that Cyg X-1  and GX~339--4 span the  same range in
hard  colour (although  note that  GX~339--4 soft-state  data are  not
included  in this  plot).  The  overall shape  drawn  by  Cyg X-1  and
GX~339--4 is similar: As the source softens, the lags increase until a
maximum is  reached, and the lags  decrease. In Cyg X-1  the turn over
happens at softer colours than in  GX~339--4, and the decrease in lags
continue to  much lower  values than in  GX~339-4. The  differences in
absolute value of  phase-lag (and that of the hard  color at which the
lags change sharply)  could be related to differences  in the system's
characteristics (e.g.  black-hole mass)  or to the  fact that  Cyg X-1
does  not reach  the  typical  soft state  seen  in  other BH  systems
\citep[e.g.,][and  references therein]{Grinberg14}.  In any  case, the
fact that Cyg X-1 shows the turn-over as it transits to a softer state
while   it   does    not   show   any   QPOs    \citep[see   ][for   a
  discussion]{Grinberg14} supports  our conclusion that  the turn-over
in phase  lags observed in  GX~339--4 is  not directly related  to the
appearance or  disappearance of QPOs  (see Section~\ref{sec:results}).
It remains to be seen whether the  lags also drop in GX~339--4 when it
reaches the soft state.

\begin{figure*} 
\centering
\resizebox{2\columnwidth}{!}{\rotatebox{-90}{\includegraphics[clip]{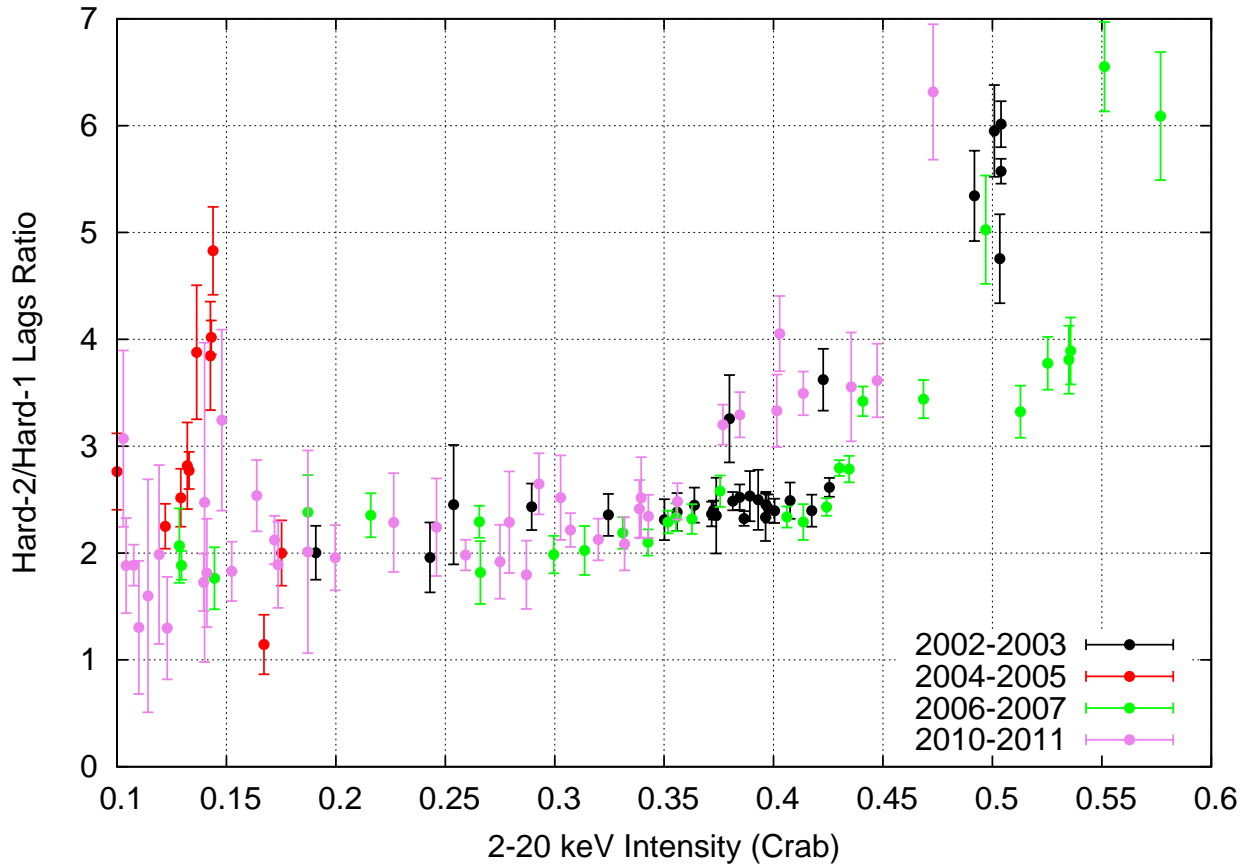}}}
\caption{The ratio between the lags in the H2 ($\simeq16-35$ keV) and
  H1 ($\simeq5.7-15$ keV) bands as a function of Intensity (in units
  of the Crab Nebula). Lags correspond to the average phase lag
  divided by $2\pi$ in the $0.008-5$ Hz frequency range using as
  reference the soft band ($\simeq2-5.7$ keV). We only show the ratios
  for intensities higher than 0.1 Crab, as below that the lags in
  single observations are not well constrained (however, the average
  ratio between lags is still $\sim$2). }
\label{fig:ratio}
\end{figure*}


\subsection{On the origin of the lags}\label{sec:origin}

In GX 339--4 the photons in both hard bands, H1 and H2, lag the
photons in the soft band.  This is consistent with previous results of
the lags in the LHS of this and other BHC
\citep[e.g.,][]{Vanderklis87, Nowak99, Belloni05, Hua97, Kazanas97,
  Cui00, Munoz11a, Grinberg14} and NS \citep{Ford99} LMXBs .

Hard lags have been originally explained as due to inverse Compton
scattering of soft disc photons in a uniform cloud of hot electrons
(the so-called corona) close to the compact object
\citep[e.g.,][]{Payne80, Miyamoto88, Ford99, Nowak99}.
With each scattering the energy of a photon ($E$) increases
approximately by $\Delta E/E = kT_e/(m_e c^2)$ \citep[e.g.,
][]{Miyamoto88, Nowak99}, where $kT_e$ is the temperature of the
electrons in the corona. This implies that the longer the path length
of the photon has to travel before leaving the corona, the longer the
time lag, and the higher the energy of the photon.
In this scenario the minimum time lag between hard and soft X-rays
would be of the order of the light crossing time of the Comptonising
region \citep[e.g.,][]{Miyamoto89, Kazanas97, Nowak99}, while the
maximum time lag must be less than the size of the emitting region
divided by the slowest propagation speed (sound or thermal wave speed)
in that region \citep[e.g][]{Nowak99}.
Very roughly, the average phase-lags in GX~339--4 would imply a
light-crossing time of $\sim$21000 km, or $\sim$700 Schwarzschild
radii ($R_S$) for a 10 $M_\odot$ black hole \citep[or much larger
  radii if one only takes the longest lags observed, see, e.g.,
][]{Kazanas97,Hua97} .

The main obstacle to this scenario \citep[e.g.][]{Nowak99,Lin00,
  Maccarone00, Poutanen01} is that in a constant-density corona the
energy required to maintain a population of highly energetic electrons
is only available very close to the black hole, whereas the magnitude
of the lags imply either very energetic electrons far away from the
black hole, or a corona with an optical depth much larger than deduced
from spectral fits \citep{Wardzinski02, Miyakawa08}.
The problem becomes worse if, to explain the frequency dependence of
the time lags in these systems, the density of the corona goes as
$r^{-1}$ \citep[i.e. inversely proportional to the distance to the
  compact object, see, e.g.,][]{Kazanas97,Hua97}. In addition,
\citet{Maccarone00} showed that the energy dependence of the width of
the autocorrelation function of the X-ray light curves produced by
Compton scattering in an isotropic corona is inconsistent with the
observations.

These difficulties lent support to the idea of \citet{Lyubarskii97}
that the variability (and the lags) in the light curve of these
systems could be due to small variations of the viscosity parameter in
the accretion disc.  These variations render fluctuations of the mass
accretion rate that propagate through the disc, and eventually
dissipate in the corona. Several authors explored the effect of this
idea on the lag spectrum of accreting sources \citep[e.g.,][]{Misra00,
  Kotov01, Arevalo06, Uttley11}. \citet{Arevalo06} carried out a
detailed study of the dependence of the power density and lag spectrum
\citep[and coherence function;][]{Vaughan97} upon, among others, the
emissivity profile of the disc, and the geometry and the viscosity
parameter of the accretion flow. Using this model, \citet{Arevalo06}
were able to reproduce the observed lag spectra in an {\em RXTE}
observation of Cyg X-1 and an XMM-Newton observation of the narrow
line Seyfert 1 galaxy NGC 4051. However, while the same model can
reproduce the $2-3$ vs. $0.5-0.9$ keV lags observed with {\em
  XMM-Newton} in GX 339--4 in the low-hard state during the 2004
outburst, it fails to explain the $6-9$ vs. $2-3$ keV in the same
observation \citep{Uttley11}.

\citet{Giannios04}, based on the work of \citet{Reig03} and
\citet{Kylafis08}, proposed that the hard lags could be due to
Comptonisation of soft disc photons in the jet that is present in the
low-hard state of galactic BHCs.
The proposed mechanism is the same as in the case of Compton
scattering in a corona (see above), but in this case the anisotropy of
the scattering process along the jet resolves the energy problem
compared to the case of the lags produced in a (spherical)
corona\footnote{\citet{Uttley11} argued that, contrary to what they
  observed in GX 339--4, this model predicts that the $2-3$ keV vs.
  $0.5-0.9$ keV lags should be smaller than the $6-9$ keV vs. $2-3$
  keV lags due to dilution by the direct seed photons from the disc at
  soft energies. However, \citet{Kara13} recently showed that
  contamination by either Poisson noise or an incoherent component
  does not affect the lags of a coherent signal. The findings of
  \citet{Uttley11} would still be consistent with the jet model of the
  lags if, for instance, the fluctuations in the disc were Poissonian,
  while the observed variability was due to the response (Green)
  function of the Comptonising component (the jet in this case).}.
Furthermore, in this model the width of the autocorrelation function
decreases with energy, in agreement with observations
\citep{Maccarone00}.
It is interesting that the radio emission from the jet is quenched
when the source moves from the low-hard to the soft-intermediate state
\citep[e.g.,][and references therein]{Fender09}, while coincidentally,
in GX 339--4 we find that the lags drop as the source moves into the
hard-intermediate and the soft-intermediate state.
%


The sudden change we observed in GX 339--4 is similar to that seen in
Cyg X-1, for which the magnitude of the lags drop abruptly as the
source moves out of the hard state (see Figure~\ref{fig:Comp}). In the
model where the lags are produced by Comptonisation in the jet, the
increase of the lags as the intensity of the source increases in the
LHS of GX 339--4 (see Figure~\ref{fig:Lags}) would be due to an
increase of the height and/or the radius of the base of the jet
\citep[see figure 5 in][]{Giannios04}. On the other hand, the drop of
the lags could indicate that the optical depth of the Comptonising
region in the jet drops, consistent with the interpretation that the
spectrum of the radio jet changes from optically thick to optically
thin synchrotron emission at this point in an outburst \citep[see,
  e.g.,][]{Fender09,Corbel13}.

Interestingly, the maximum of the lags in GX 339--4 coincides, within
$\pm 1$ day, with the time at which there is a sudden increase of the
cut-off energy $E_{cut}$ of the cut-off power law that \citet{Motta09}
fitted to the energy spectrum of GX 339--4 during the 2006-2007
outburst (see their figure 6).
Furthermore, from figure 6 in \citet{Motta09} it is also apparent that
the $E_{cut}$ of the hard spectral component in GX 339$-$4 remained
more or less constant for the first $\sim 20$ days of the 2006-20076
outburst, after which, $E_{cut}$ started to decrease.
From our Figure \ref{fig:HIDPP} it appears that the behaviour of the
phase lags also changes around that time; at the beginning of the
outburst the phase lags appear to be constant, and when the intensity
reaches $\sim 0.2$ Crab the lags start to increase .

To further investigate this, in Figure~\ref{fig:cutoff} we plot the
temporal evolution of $E_{cut}$ \citep[data from][]{Motta09} and the
phase lags for the first 45 days of the 2006-2007 outburst.
This Figure shows that both $E_{cut}$ and the phase lag remain more or
less constant until approximately MJD 50952, date when $E_{cut}$
starts to decrease while the phase lags start to increase. Under the
assumption that both quantities are correlated, we fitted a broken
power law to both curves simultaneously, with the condition that the
break was the same for both fits; we further assumed that the
power-law index before the break was zero, i.e. both $E_{cut}$ and the
phase lags are constant before the break. The continuous line in both
panels of Figure~\ref{fig:cutoff} represents the best-fitting model
($\chi^2/dof=4.1$ for 30 dof) with a break at $MJD-50946=20.1\pm1$
day. The fit improves ($\chi^2/dof=2.6$ for 28 dof) if we let the
power-law indices before the break free to vary during the fits; in
this case the break happens at $MJD-50946 = 17.4\pm0.7$ day.
None of the fits is statistically acceptable and there is a priori no
reason to assume that a broken power-law is the correct model to
describe the data. Even if our proposed model was correct, the lack of
data at the break precludes us from concluding whether the break
occurs simultaneously in both curves.  While this needs to be
investigated in more detail (e.g., with data from other outbursts),
Figure~\ref{fig:cutoff} strongly suggests that, already at this early
phase of the outburst, both $E_{cut}$ and the phase lags are
correlated.
This result, together with the fact that the maximum lag we measured
in GX~339--4 coincides (within $\pm 1$ day) with the time at which
there is another (sudden) increase of $E_{cut}$, suggests that the
average magnitude of the lags are related to the properties of
corona/jet \citep[e.g.,][]{Giannios04}, rather than to the disc
\citep[e.g.,][]{Lyubarskii97,Uttley11}.

Intriguingly, there is no apparent discontinuity in the time evolution
of $E_{cut}$ at the point at which the lags start to increase very
rapidly, e.g. the green point during the 2006-2007 outburst in the
left panel of Figure~\ref{fig:HIDPP}. This point, however, coincides
with the turn-the-corner point in the HID (right panel of
Figure~\ref{fig:HIDPP}; see Figure~\ref{fig:Lags} for the other
outbursts), which defines the transition from the hard to the
intermediate state. To investigate this in more detail, in
Figure~\ref{fig:ratio} we plot the ratio between the 0.008--5 Hz
averaged lags in GX 339--4 measured in the H2 and H1 bands as a
function of the intensity of the source.
At intensities lower than about 0.35 Crab (for the weak 2004-2005
outburst this value is $\sim 0.12$ Crab) the ratio of the lags remains
approximately constant at a value of $\sim2$, and as the intensity
rises further the ratio increases abruptly up to values of $6-7$.
The ratio in Figure~\ref{fig:ratio} deviates from a constant value at
the same time that the source reaches the ``turn the corner'' point in
the HID (Figure~\ref{fig:Lags}), implying that the transition to the
intermediate state has an effect on the characteristic of the
component in the accretion flow that produces the lags, e.g., the
geometry of the jet in the model of \citet{Reig03} and
\citet{Giannios04}, or the accretion disc in the model of
\citet{Lyubarskii97} and \citet{Arevalo06}. \citet{Motta09} found
that, at least in the 2006-2007 outbursts of GX~339--4, the accretion
disc does not contribute significantly to the emission in the RXTE/PCA
band in this phase of the outburst \citep[][used the same data that we
  used in our paper for the study of the lags]{Motta09}. Together with
our findings, this implies that, even if the lags are due to the
propagation of fluctuations in a relatively cool disc that peaks
outside the RXE/PCA band, Comptonisation would still be required to
explain lags.

%
%
%
%

\section{Summary}

We studied the evolution of the X-ray lags of the transient BHC-LMXB
GX~339--4 as a function of the position of the source in the HID.
In the low-hard state the lags correlate with X-ray intensity,
and as the source starts its transition to the intermediate/soft states,
the lags first increase faster, and then appear to reach a maximum,
although the exact evolution depends on the outburst and the energy
band used to calculate the lags.
Changes in the lags appear to coincide with changes in Optical/NIR
flux, fractional RMS amplitude, and probably the appearance of a
thermal component in the X-ray spectra, strongly suggesting that lags
can be very useful to understand the physical changes that GX~339--4
undergoes during an outburst. We find evidence for a connection
between the evolution of the cut-off energy $E_{cut}$ (of the energy
spectra) and the phase lags, suggesting that the average magnitude of
the lags are related to the properties of corona/jet, rather than to
the disc.

The behaviour of the lags in GX~339--4 is similar to that of the BH
Cygnus~X--1 \citep{Grinberg14}, suggesting similar phenomena could be
observable in other BH systems.
Understanding how lags evolve in other BH outbursts, and whether the
correlation with intensity, maximum value at $\sim$20\% fractional RMS
amplitude, etc., is common to all BHC outbursts, or whether that
depends on the characteristics of the system (e.g., neutron star
vs. black hole, BHC mass, spin, inclination of the orbit with respect
to the observer, radio flux, etc), is a necessary input for current
and future models that describe both X-ray variability and
multi-wavelength observations.

\section*{Acknowledgments}
We thanks N. Kylafis, T. Belloni, P. Uttley, I. McHardy and
D. Emmanoulopoulos for insightful discussions. We also thank the
referee for very useful suggestions. DA acknowledges support from the
Royal Society. This research has made use of data and/or software
provided by the High Energy Astrophysics Science Archive Research
Center (HEASARC), which is a service of the Astrophysics Science
Division at NASA/GSFC and the High Energy Astrophysics Division of the
Smithsonian Astrophysical Observatory.This research has made use of
NASA's Astrophysics Data System.


\label{lastpage}
\end{document}